\documentclass[11pt,a4paper]{article}

\usepackage[cp1251]{inputenc}
\usepackage[T2A]{fontenc}
\usepackage[english,russian]{babel}
\usepackage[dvips]{graphicx}
\usepackage[tbtags]{amsmath}
\usepackage{amsfonts,amssymb,mathrsfs,amscd}

\usepackage{amsxtra}
\usepackage{amsthm}
\usepackage{latexsym}
\usepackage{array}
\usepackage{multirow}
\usepackage{hhline}

\usepackage{latexsym}

\newtheorem{propos}{Предложение}

\numberwithin{equation}{section}

\newcounter{myta}

\newcommand {\mP}{\mathcal{P}}

\newcommand {\ds}{\displaystyle}
\newcommand {\mm} {\mathcal{M}_1}
\newcommand {\mn} {\mathcal{M}_2}
\newcommand {\mo} {\mathcal{M}_3}
\newcommand {\ml} {\mathcal{M}_4}

\begin{document}
\begin{flushright}
{\it 13.02.2013}
\end{flushright}

\begin{center}
\textbf{\Large Phase topology of one nonclassical integrable problem
of dynamics\\[5mm]
P.\,E.~Ryabov}\\[3mm]
\small

Financial University under the Government of the Russian Federation, Moscow, Russia

E-mail: orelryabov@mail.ru
\end{center}

\begin{center}
\textbf{Abstract}
\end{center}
\normalsize

 {\footnotesize We consider the integrable system with three degrees of
freedom for which Sokolov and Tsiganov specified Lax represen\-tation. Lax representation
generalizes $L$\,--$A$ pair of the Kowalevski gyrostat in two constant fields, found by
A.\,G.~Reyman and M.\,A.~Semenov-Tian-Shansky. In the paper, we give the explicit
formulas for the (independent almost everywhere) additional first integrals $K$ and $G$.
These integrals are functionally connected with factors of a spectral curve of $L$\,--$A$
pair by Sokolov and Tsiganov. Due to this form of additional integrals $K$ and $G$,
without constant gyrostatic moment, we managed to find analytically two invariant
four-dimensional submanifolds on which the induced dynamic system is almost everywhere
Hamiltonian system with two degrees of freedom. System of equations that describes one of
these invariant submanifolds is a generalization of invariant relations of an integrable
Bogoyavlensky case in dynamics of a rigid body. To describe phase topology of the system
as a whole we use the method of critical subsystems. For each subsystem, we construct the
bifurcation diagrams and specify the bifurcations of Liouville tori both in subsystems,
and in the system as a whole.

Bibliography:  24 titles.

\vspace{3mm} Keywords: the Kirchhoff equations, a completely integrable Hamiltonian
systems, the spectral curve, the momentum map, the bifurcation diagram, the bifurcations
of Liouville tori.

Mathematical Subject Classification 2000: 70E17, 70G40 }

\vspace{20mm}

\begin{center}
\textbf{\Large Фазовая топология одной неклассической интегрируемой\\ задачи
динамики\\[5mm]
П.\,Е.~Рябов}\\[3mm]

\small

Финансовый университет при Правительстве Российской Федерации

E-mail: orelryabov@mail.ru
\end{center}

\begin{center}
\textbf{Аннотация}
\end{center}

{\footnotesize Рассматривается интегрируемая система с тремя степенями свободы, для
которой Соколовым и Цыгановым указано представление Лакса. Представление Лакса обобщает
$L$\,--$A$ пару для гиростата Ковалевской в двойном поле, найденную А.\,Г.~Рейманом и
M.\,A.~Семеновым-Тян-Шанским. В данной работе мы приводим явные формулы для (независимых
почти всюду) дополнительных первых интегралов $K$ и $G$, которые функционально связаны с
коэффициентами спектральной кривой $L$\,--$A$ пары Соколова и Цыганова. Благодаря такой
форме дополнительных интегралов $K$ и $G$ при отсутствии постоянного гиростатического
момента, удалось выделить аналитически два инвариантных четырехмерных подмногообразия, на
которых индуцированная динамическая система является почти всюду гамильтоновой с двумя
степенями свободы. Система уравнений, задающая одно из инвариантных подмногообразий,
является обобщением инвариантных соотношений интегрируемого случая Богоявленского в
динамике твердого тела. Для описания фазовой топологии всей системы в целом используется
метод критических подсистем. Для каждой подсистемы построены бифуркационные диаграммы и
указаны бифуркации торов Лиувилля как внутри подсистем, так и во всей системе в целом.

Библиография: 24 назв.

\vspace{3mm} Ключевые слова: уравнения Кирхгофа, вполне интегрируемые гамильтоновы
системы, спектральная кривая, отображение момента, бифуркационная диаграмма, бифуркации
торов Лиувилля }

\vspace{3mm}

\section{Введение}

Прошло одиннадцать лет с тех пор, как в \cite{sok_tsig_2002} была доказана
интегрируемость системы  уравнений Кирхгофа

\begin{equation}\label{eq_1_0}
\begin{array}{l}
 \ds{\dot{\boldsymbol M}={\boldsymbol M}\times\frac{\partial H}{\partial{\boldsymbol
M}}+ {\boldsymbol\alpha}\times\frac{\partial H}{\partial{\boldsymbol\alpha}}+
{\boldsymbol\beta}\times\frac{\partial H}{\partial{\boldsymbol\beta}},}\\[5mm]
\ds{\dot{\boldsymbol \alpha}={\boldsymbol\alpha}\times\frac{\partial
H}{\partial{\boldsymbol M}},\quad \dot{\boldsymbol
\beta}={\boldsymbol\beta}\times\frac{\partial H}{\partial{\boldsymbol M}}}
\end{array}
\end{equation}
c гамильтонианом
\begin{equation}\label{eq_1_1}
H_1=M_1^2+M_2^2+2M_3^2+2\lambda
M_3-2(\alpha_1+\beta_2)+2\varepsilon_1(M_2\alpha_3-M_3\alpha_2+M_3\beta_1-M_1\beta_3)
\end{equation}
Здесь трехмерные векторы ${\boldsymbol M}, {\boldsymbol\alpha}, {\boldsymbol\beta}$
представляют собой проекции "импульсивного момента"\ и двух силовых полей ("импульсивных
сил") на оси, жестко связанные с твердым телом, $\varepsilon_1$ -- параметр деформации.

 Соответствующая скобка Ли--Пуассона задается формулами
\begin{equation}\label{eq_1_2}
\begin{array}{l}\{M_i,M_j\}=\varepsilon_{ijk}M_k, \{M_i,\alpha_j\}=\varepsilon_{ijk}\alpha_k,
\{M_i,\beta_j\}=\varepsilon_{ijk}\beta_k,\\[5mm]
\{\alpha_i,\alpha_j\}=0, \{\alpha_i,\beta_j\}=0, \{\beta_i,\beta_j\}=0, \\[5mm]
\varepsilon_{ijk}=\frac{1}{2}(i-j)(j-k)(k-i),\quad 1\leqslant i,j,k\leqslant 3.
\end{array}
\end{equation}

Функциями Казимира являются выражения ${\boldsymbol\alpha}^2$,
${\boldsymbol\alpha}\cdot{\boldsymbol\beta}$ и ${\boldsymbol\beta}^2$.

 Относительно
скобки Ли--Пуассона, заданной соотношениями (\ref{eq_1_2}), систему (\ref{eq_1_0}) можно
представить в гамильтоновом виде:
\begin{equation*}
\dot x=\{H_1,x\},
\end{equation*}
где через $x$ обозначена любая из координат.

 В работе \cite{sok_tsig_2002} указана
соответствующая $L-A$ пара для гамильтониана (\ref{eq_1_1}). Хорошо известно, что
коэффициенты спектральной кривой ${\cal E}_1(z,\zeta)=0$ для $L-A$ пары всегда являются
первыми интегралами. Оказывается необходимый дополнительный интеграл (коэффициент при
$z^4$ в алгебраической кривой ${\cal E}_1(z,\zeta)=0$) можно выразить через другие
(независимые почти всюду) дополнительные интегралы той же системы (формула
(\ref{eq_1_7})). Деформации интегрируемых гамильтонианов, анонсированные в
\cite{sok_tsig_2002}, упоминаются в книгах \cite{bormam2003} (\cite[формула (4.18),
c.~128]{bormam2003}) и \cite{bormam2005} (\cite[замечание~2, c.~265]{bormam2005}). При
отсутствии второго силового поля ($\boldsymbol\beta=\boldsymbol 0$) и наличии ненулевого
параметра $\lambda$ (параметра гиростатического момента) интегрируемость доказана
В.\,В.~Соколовым. Явное выражение дополнительного интеграла (на алгебре $e(3)$)
содержится в работах \cite{Sok02}, \cite{Sok03}. В \cite{ryab_mtt2007} дополнительный
интеграл Соколова представлен в виде, удобном для исследования фазовой топологии в
системе с двумя степенями свободы.

Для гамильтониана (\ref{eq_1_1}) дополнительные интегралы имеют следующий вид:
\begin{equation}\label{eq_1_4}
\begin{array}{l}
\ds{K_1=\left[\frac{1}{2}(M_1^2-M_2^2)+\alpha_1-\beta_2+\varepsilon_1(M_3\alpha_2-M_2\alpha_3+M_3\beta_1-M_1\beta_3)-\frac{1}{2}\varepsilon_1^2({\boldsymbol\alpha}^2-{\boldsymbol\beta}^2)\right]^2+}\\[3mm]
\ds{+[M_1M_2+\alpha_2+\beta_1+\varepsilon_1(M_1\alpha_3-M_3\alpha_1+M_3\beta_2-M_2\beta_3)]^2-}\\[3mm]
\ds{-\lambda[(M_3+\lambda)(M_1^2+M_2^2)+2(\alpha_3M_1+\beta_3M_2)]+\lambda\varepsilon_1^2({\boldsymbol\alpha}^2+{\boldsymbol\beta}^2)M_3+}\\[3mm]
\ds{+2\lambda\varepsilon_1[\alpha_2M_1^2-\beta_1M_2^2-(\alpha_1-\beta_2)M_1M_2]-2\lambda\varepsilon_1^2\omega_\gamma-}\\[3mm]
\ds{-2\varepsilon_1^2({\boldsymbol\alpha}\cdot{\boldsymbol\beta})[\alpha_2+\beta_1+M_1M_2+\varepsilon_1(\alpha_3M_1-\alpha_1M_3+\beta_2M_3-\beta_3M_2)],}\\[3mm]

\ds{G_1=\omega_\alpha^2+\omega_\beta^2+2(M_3+\lambda)\omega_\gamma-2{\boldsymbol\alpha}^2\beta_2-2{\boldsymbol\beta}^2\alpha_1+}\\[3mm]
\ds{+2\varepsilon_1[{\boldsymbol\alpha}^2(M_3\beta_1-M_1\beta_3)+{\boldsymbol\beta}^2(M_2\alpha_3-M_3\alpha_2)]+}\\[3mm]
\ds{+2({\boldsymbol\alpha}\cdot{\boldsymbol\beta})[\alpha_2+\beta_1+\varepsilon_1(\alpha_3M_1-\alpha_1M_3+\beta_2M_3-\beta_3M_2)],}
\end{array}
\end{equation}
где
\begin{equation*}
\begin{array}{l}
\omega_\alpha=M_1\alpha_1+M_2\alpha_2+M_3\alpha_3,\\
\omega_\beta=M_1\beta_1+M_2\beta_2+M_3\beta_3,\\
\omega_\gamma=M_1(\alpha_2\beta_3-\alpha_3\beta_2)+M_2(\alpha_3\beta_1-\alpha_1\beta_3)+
M_3(\alpha_1\beta_2-\alpha_2\beta_1).
\end{array}
\end{equation*}

Функции $K_1$ и $G_1$ записаны таким образом, чтобы их можно было сравнить с интегралами
$I_1$ и $I_2$ из работы \cite{ReySem1987} или \cite{BobReySem1989}. А именно, если
положить значение параметра деформации $\varepsilon_1$ равным нулю, то получаются
выражения для $I_1$ и $I_2$ (\cite[формула~(5), c.~57]{ReySem1987}).

Укажем явное выражение алгебраической кривой ${\cal E}_1(z,\zeta)$:

\begin{equation*}\label{eq_1_5}
{\cal E}_1(z,\zeta)\,:\, \, d_4^1\zeta^4+d_2^1\zeta^2+d_0^1=0,
\end{equation*}
где
\begin{equation*}\label{eq_1_6}
\begin{array}{l}
\ds{d_4^1=-z^4-\varepsilon_1^2({\boldsymbol\alpha}^2+{\boldsymbol\beta}^2)z^2-\varepsilon_1^4[{\boldsymbol\alpha}^2{\boldsymbol\beta}^2-({\boldsymbol\alpha}\cdot{\boldsymbol\beta})^2],}\\[3mm]
\ds{d_2^1=2z^6+[\varepsilon_1^2({\boldsymbol\alpha}^2+{\boldsymbol\beta}^2)-h_1-\lambda^2]z^4
+({\boldsymbol\alpha}^2+{\boldsymbol\beta}^2-\varepsilon_1^2g_1)z^2+2\varepsilon_1^2[{\boldsymbol\alpha}^2{\boldsymbol\beta}^2-({\boldsymbol\alpha}\cdot{\boldsymbol\beta})^2],}\\[3mm]
\ds{d_0^1=-z^8+h_1z^6+d_{04}^1z^4+g_1z^2-[{\boldsymbol\alpha}^2{\boldsymbol\beta}^2-({\boldsymbol\alpha}\cdot{\boldsymbol\beta})^2].}
\end{array}
\end{equation*}
Коэффициент $d_{04}^1$ при $z^4$, как отмечалось выше, всегда является первым интегралом.
Нам удалось выразить коэффициент $d_{04}^1$  через казимиры ${\boldsymbol\alpha}^2$,
${\boldsymbol\beta}^2$ и другие (независимые почти всюду) дополнительные интегралы
(\ref{eq_1_4}) той же системы:
\begin{equation}\label{eq_1_7}
\begin{array}{l}
\ds{d_{04}^1=\varepsilon_1^2g_1+k_1-\frac{1}{4}[h_1^2+2\varepsilon_1^2({\boldsymbol\alpha}^2+{\boldsymbol\beta}^2)h_1
+\varepsilon_1^4({\boldsymbol\alpha}^2-{\boldsymbol\beta}^2)^2]-{\boldsymbol\alpha}^2-{\boldsymbol\beta}^2,}
\end{array}
\end{equation}
где $h_1$, $g_1$ и $k_1$ -- постоянные первых интегралов (\ref{eq_1_1}) и (\ref{eq_1_4}).

Для полноты приведем формулы для другой деформации гамильтониана $H_2$, интегрируемость
которой также доказана в \cite{sok_tsig_2002}:

\begin{equation}\label{eq_1_8}
H_2=M_1^2+M_2^2+2M_3^2+2\lambda
M_3-2\varepsilon_2(\alpha_1+\beta_2)+2(M_2\alpha_3-M_3\alpha_2+M_3\beta_1-M_1\beta_3).
\end{equation}
Здесь $\varepsilon_2$ -- параметр деформации.

Дополнительные интегралы $K_2$ и $G_2$ для гамильтониана (\ref{eq_1_8}) имеют
соответственно вид:

\begin{equation}\label{eq_1_9}
\begin{array}{l}
\ds{K_2=\left[\frac{1}{2}(M_1^2-M_2^2)+\varepsilon_2(\alpha_1-\beta_2)+M_3\alpha_2-M_2\alpha_3+M_3\beta_1-M_1\beta_3-\frac{1}{2}({\boldsymbol\alpha}^2-{\boldsymbol\beta}^2)\right]^2+}\\[3mm]
\ds{+[M_1M_2+\varepsilon_2(\alpha_2+\beta_1)+M_1\alpha_3-M_3\alpha_1+M_3\beta_2-M_2\beta_3]^2-}\\[3mm]
\ds{-\lambda[(M_3+\lambda)(M_1^2+M_2^2)+2\varepsilon_2(\alpha_3M_1+\beta_3M_2)]+\lambda({\boldsymbol\alpha}^2+{\boldsymbol\beta}^2)M_3+}\\[3mm]
\ds{+2\lambda[\alpha_2M_1^2-\beta_1M_2^2-(\alpha_1-\beta_2)M_1M_2]-2\lambda\omega_\gamma-}\\[3mm]
\ds{-2({\boldsymbol\alpha}\cdot{\boldsymbol\beta})[\varepsilon_2(\alpha_2+\beta_1)+M_1M_2+\alpha_3M_1-\alpha_1M_3+\beta_2M_3-\beta_3M_2],}\\[3mm]

\ds{G_2=\omega_\alpha^2+\omega_\beta^2+2(M_3+\lambda)\omega_\gamma-2\varepsilon_2({\boldsymbol\alpha}^2\beta_2+{\boldsymbol\beta}^2\alpha_1)+}\\[3mm]
\ds{+2[{\boldsymbol\alpha}^2(M_3\beta_1-M_1\beta_3)+{\boldsymbol\beta}^2(M_2\alpha_3-M_3\alpha_2)]+}\\[3mm]
\ds{+2({\boldsymbol\alpha}\cdot{\boldsymbol\beta})[\varepsilon_2(\alpha_2+\beta_1)+\alpha_3M_1-\alpha_1M_3+\beta_2M_3-\beta_3M_2],}
\end{array}
\end{equation}

Алгебраическая кривая ${\cal E}_2(z,\zeta)$ задается выражением:
\begin{equation*}\label{eq_1_10}
{\cal E}_2(z,\zeta)\,:\, \, d_4^2\zeta^4+d_2^2\zeta^2+d_0^2=0,
\end{equation*}
где
\begin{equation*}
\begin{array}{l}
\ds{d_4^2=-z^4-({\boldsymbol\alpha}^2+{\boldsymbol\beta}^2)z^2-[{\boldsymbol\alpha}^2{\boldsymbol\beta}^2-({\boldsymbol\alpha}\cdot{\boldsymbol\beta})^2],}\\[3mm]
\ds{d_2^2=2z^6+({\boldsymbol\alpha}^2+{\boldsymbol\beta}^2-h_2-\lambda^2)z^4
+[\varepsilon_2^2({\boldsymbol\alpha}^2+{\boldsymbol\beta}^2)-g_2]z^2+2\varepsilon_2^2[{\boldsymbol\alpha}^2{\boldsymbol\beta}^2-({\boldsymbol\alpha}\cdot{\boldsymbol\beta})^2],}\\[3mm]
\ds{d_0^2=-z^8+h_2z^6+}\\[3mm]
\ds{+[g_2+k_2-\frac{1}{4}[h_2^2+2({\boldsymbol\alpha}^2+{\boldsymbol\beta}^2)h_2
+({\boldsymbol\alpha}^2-{\boldsymbol\beta}^2)^2]-\varepsilon_2^2({\boldsymbol\alpha}^2+{\boldsymbol\beta}^2)]z^4+}\\[3mm]
\ds{+\varepsilon_2^2g_2z^2-\varepsilon_2^4[{\boldsymbol\alpha}^2{\boldsymbol\beta}^2-({\boldsymbol\alpha}\cdot{\boldsymbol\beta})^2].}
\end{array}
\end{equation*}
Здесь через $h_2$, $k_2$, $g_2$ обозначены постоянные первых интегралов (\ref{eq_1_8}) и
(\ref{eq_1_9}).

Как показано в \cite{Kh34}, \cite{KhNd2007}, \cite{Khmath2009} без ограничения общности
можно считать векторы $\boldsymbol\alpha$ и $\boldsymbol\beta$ взаимно ортогональными,
причем $|\boldsymbol\alpha|\geqslant|\boldsymbol\beta|$. Тогда геометрические интегралы,
порождаемые функциями Казимира, запишутся в виде
\begin{equation}\label{eq_1_12}
|\boldsymbol\alpha|^2=a^2,\quad |\boldsymbol\beta|^2=b^2,\quad
{\boldsymbol\alpha}\cdot{\boldsymbol\beta}=0, \quad (a\geqslant b>0).
\end{equation}

Цель настоящей работы --- предъявить новые  инвариантные четырехмерные подмногообразия,
на которых индуцированная динамическая система является почти всюду гамильтоновой с двумя
степенями свободы.  Мы также анонсируем атлас бифуркационных диаграмм и сами диаграммы
таких систем. Мы приводим новый сетевой инвариант на изоэнергетической поверхности,
которого нет в списке сетевых инвариантов в задаче о движении волчка Ковалевской в
двойном поле сил.

\section{Первая система -- обобщение интегрируемого случая Богоявленского в динамике твердого тела}

При отсутствии постоянного гиростатического момента (параметр $\lambda$ равен нулю)
дополнительные интегралы (\ref{eq_1_4}) принимают вид:
\begin{equation}\label{eq_2_1}
\begin{array}{l}
\ds{K=Z_1^2+Z_2^2,}\\[3mm]
\ds{G=\omega_\alpha^2+\omega_\beta^2+2M_3\omega_\gamma-2{\boldsymbol\alpha}^2\beta_2-2{\boldsymbol\beta}^2\alpha_1+}\\[3mm]
\ds{+2\varepsilon[{\boldsymbol\alpha}^2(M_3\beta_1-M_1\beta_3)+{\boldsymbol\beta}^2(M_2\alpha_3-M_3\alpha_2)],}\\[3mm]
\end{array}
\end{equation}
где
\begin{equation*}\label{eq_2_2}
\begin{array}{l}
\ds{Z_1=\frac{1}{2}(M_1^2-M_2^2)+\alpha_1-\beta_2+\varepsilon(M_3\alpha_2-M_2\alpha_3+M_3\beta_1-M_1\beta_3)-\frac{1}{2}\varepsilon^2({\boldsymbol\alpha}^2-{\boldsymbol\beta}^2),}\\[3mm]
\ds{Z_2=M_1M_2+\alpha_2+\beta_1+\varepsilon(M_1\alpha_3-M_3\alpha_1+M_3\beta_2-M_2\beta_3)}.
\end{array}
\end{equation*}
Здесь и далее для краткости обозначено $\varepsilon=\varepsilon_1$.

Рассмотрим фазовое пространство $\mP^6$, задаваемое уравнениями (\ref{eq_1_12}).

\begin{propos}
Система соотношений
\begin{equation}\label{eq_2_3}
Z_1=0,\quad Z_2=0
\end{equation}
определяет инвариантное четырехмерное подмногообразие $\mm$ фазового пространства $\mP^6$
уравнений (\ref{eq_1_0}) с гамильтонианом
\begin{equation}\label{eq_2_4}
H=M_1^2+M_2^2+2M_3^2-2(\alpha_1+\beta_2)+2\varepsilon(M_2\alpha_3-M_3\alpha_2+M_3\beta_1-M_1\beta_3)
\end{equation}
\end{propos}
\begin{proof}
Действительно,  с учетом (\ref{eq_1_12}) и (\ref{eq_2_3}) имеем:
\begin{equation*}\label{eq_2_5}
\dot Z_1=\{H,Z_1\}=4M_3Z_2=0,\quad \dot Z_2=\{H,Z_2\}=-4M_3Z_1=0.
\end{equation*}
\end{proof}

\begin{propos}
Функция
\begin{equation*}\label{eq_2_6}
\begin{array}{l}
\ds{F_0=\{Z_1,Z_2\}=}\\[3mm]
\ds{=[M_1^2+M_2^2+2\varepsilon^2(\alpha_1\beta_2-\alpha_2\beta_1)-\varepsilon^2(a^2+b^2)]M_3+2(M_1\alpha_3+M_2\beta_3)+}\\[3mm]
\ds{+2\varepsilon^2[M_1(\alpha_2\beta_3-\alpha_3\beta_2)-M_2(\alpha_1\beta_3-\alpha_3\beta_1)]+
2\varepsilon[M_1M_2(\alpha_1-\beta_2)-M_1^2\alpha_2+M_2^2\beta_1]}
\end{array}
\end{equation*}
является первым интегралом на подмногообразии $\mm$, заданном уравнениями (\ref{eq_2_3}).
\end{propos}
\begin{proof}
Действительно, в силу тождества Якоби, правила Лейбница и уравнений (\ref{eq_2_3}),
находим:
\begin{equation*}\label{eq_2_7}
\dot F_0=\{H,F_0\}=\{H,\{Z_1,Z_2\}\}=4Z_1\{M_3,Z_1\}+4Z_2\{M_3,Z_2\}=0.
\end{equation*}
\end{proof}

Отметим, что система уравнений (\ref{eq_2_3}) является обобщением инвариантных
соотношений интегрируемого случая Богоявленского в динамике твердого тела
\cite[формула~(6.6), c.~904]{Bogo}.

В точках подмногообразия $\mm$ непосредственно проверяется справедливость следующего
тождества, которое является обобщением соответствующей формулы работы \cite{Zotev2000}:
\begin{equation*}\label{eq_2_8}
\{2+\varepsilon^2[\varepsilon^2({\boldsymbol\alpha}^2+{\boldsymbol\beta}^2)+H]\}
[({\boldsymbol\alpha}^2+{\boldsymbol\beta}^2)H-2G+\varepsilon^2({\boldsymbol\alpha}^2-{\boldsymbol\beta}^2)^2]-F_0^2=0.
\end{equation*}
Перечислим формально (без условия существования) все положения равновесия
$c_i(\boldsymbol M, \boldsymbol\alpha, \boldsymbol\beta)$ и их образы $P_i$ в ${\Bbb
R}^3(h,k,g)$. Положения равновесия являются особенностями ранга $0$ отображения момента
${\cal F}=H\times K\times G$, порождаемого первыми интегралами (\ref{eq_2_1}),
(\ref{eq_2_4}). Особенности ранга $0$ образуют нульмерный остов $\Sigma_0$ бифуркационной
диаграммы $\Sigma$ (образа множества критических значений отображения момента):
\begin{equation*}\label{eq_2_9}
 \begin{array}{l}
 \ds{c_{1,2}=\left(0,\mp\frac{\sqrt{a^2\varepsilon^4-1}}{\varepsilon},0,\frac{1}{\varepsilon^2},
 0,\pm\frac{\sqrt{a^2\varepsilon^4-1}}{\varepsilon^2},0,b,0\right),}\\[5mm]
 \ds{P_{1,2}=\left(-\frac{1+a^2\varepsilon^4+2b\varepsilon^2}{\varepsilon^2},
\frac{(b\varepsilon^2-1)^4}{4\varepsilon^4},-\frac{b(b+a^2b\varepsilon^4+2a^2\varepsilon^2)}{\varepsilon^2}\right),}\\[5mm]
\ds{c_{3,4}=\left(0,\mp\frac{\sqrt{a^2\varepsilon^4-1}}{\varepsilon},0,\frac{1}{\varepsilon^2},
 0,\pm\frac{\sqrt{a^2\varepsilon^4-1}}{\varepsilon^2},0,-b,0\right),}\\[5mm]
 \ds{P_{3,4}=\left(-\frac{1+a^2\varepsilon^4-2b\varepsilon^2}{\varepsilon^2},
\frac{(b\varepsilon^2+1)^4}{4\varepsilon^4},-\frac{b(b+a^2b\varepsilon^4-2a^2\varepsilon^2)}{\varepsilon^2}\right),}\\[5mm]
\end{array}
 \end{equation*}
 \begin{equation*}
 \begin{array}{l}
\ds{c_{5,6}=\left(\pm\frac{\sqrt{b^2\varepsilon^4-1}}{\varepsilon},0,0,a,
 0,0,0,\frac{1}{\varepsilon^2},\pm\frac{\sqrt{b^2\varepsilon^4-1}}{\varepsilon^2}\right),}\\[5mm]
\ds{P_{5,6}=\left(-\frac{1+b^2\varepsilon^4+2a\varepsilon^2}{\varepsilon^2},
\frac{(a\varepsilon^2-1)^4}{4\varepsilon^4},-\frac{a(a+ab^2\varepsilon^4+2b^2\varepsilon^2)}{\varepsilon^2}\right),}\\[5mm]
\ds{c_{7,8}=\left(\pm\frac{\sqrt{b^2\varepsilon^4-1}}{\varepsilon},0,0,-a,
 0,0,0,\frac{1}{\varepsilon^2},\pm\frac{\sqrt{b^2\varepsilon^4-1}}{\varepsilon^2}\right),}\\[5mm]
\ds{P_{7,8}=\left(-\frac{1+b^2\varepsilon^4-2a\varepsilon^2}{\varepsilon^2},
\frac{(a\varepsilon^2+1)^4}{4\varepsilon^4},-\frac{a(a+ab^2\varepsilon^4-2b^2\varepsilon^2)}{\varepsilon^2}\right),}
\end{array}
 \end{equation*}
\begin{equation*}
 \begin{array}{l}
\ds{c_{9,10,11,12}=\left(\boldsymbol 0,\pm a,0,0,0,\pm b,0\right),}\\[5mm]
\ds{P_{9,10,11,12}=\left(\mp 2(a\pm b),\frac{1}{4}(a\mp b)^2[\varepsilon^2(a\pm b)-2]^2,\mp 2ab(a\pm b)\right),}\\[5mm]
\ds{c_{13-16}=\left(0,\mp\varepsilon\sqrt{a^2-b^2},\mp\frac{\sqrt{b^2\varepsilon^4-1}}{\varepsilon},
\frac{1}{\varepsilon^2},\mp\frac{\sqrt{b^2\varepsilon^4-1}}{\varepsilon^2},\pm\sqrt{a^2-b^2},
\pm\frac{\sqrt{b^2\varepsilon^4-1}}{\varepsilon^2},\frac{1}{\varepsilon^2},0\right),}\\[5mm]
\ds{P_{13-16}=\left(-\frac{\varepsilon^4(a^2+b^2)+2}{\varepsilon^2},0,-\frac{a^2+b^2+2a^2b^2\varepsilon^4}{\varepsilon^2}\right),}\\[5mm]
\end{array}
 \end{equation*}
\begin{equation*}
 \begin{array}{l}
\ds{c_{17,18}=\left(0,0,\pm\frac{\sqrt{\varepsilon^4(a+b)^2-4}}{2\varepsilon},\frac{2a}{\varepsilon^2(a+b)},
\pm\frac{\sqrt{\varepsilon^4(a+b)^2-4}a}{(a+b)\varepsilon^2},0,\mp\frac{\sqrt{\varepsilon^4(a+b)^2-4}b}{(a+b)\varepsilon^2},
\frac{2b}{\varepsilon^2(a+b)},0\right),}\\[5mm]
\ds{P_{17,18}=\left(-\frac{\varepsilon^4(a+b)^2+4}{2\varepsilon^2},0,-\frac{[\varepsilon^4(a+b)^2+4]ab}{2\varepsilon^2}\right),}\\[5mm]
\ds{c_{19,20}=\left(0,0,\pm\frac{\sqrt{\varepsilon^4(a-b)^2-4}}{2\varepsilon},\frac{2a}{\varepsilon^2(a-b)},
\pm\frac{\sqrt{\varepsilon^4(a-b)^2-4}a}{(a-b)\varepsilon^2},0,\pm\frac{\sqrt{\varepsilon^4(a-b)^2-4}b}{(a-b)\varepsilon^2},
-\frac{2b}{\varepsilon^2(a-b)},0\right),}\\[5mm]
\ds{P_{19,20}=\left(-\frac{\varepsilon^4(a-b)^2+4}{2\varepsilon^2},0,\frac{[\varepsilon^4(a-b)^2+4]ab}{2\varepsilon^2}\right).}\\[5mm]
\end{array}
 \end{equation*}

При $\varepsilon>\max\{\frac{1}{\sqrt{b}};\sqrt{\frac{2}{a-b}}\}$ подмногообразие $\mm$
содержит положения равновесия $c_{13-20}$. Наличие особенностей ранга $0$ в системе $\mm$
и их аналитическая классификация по типу позволяет применить метод круговых молекул для
анализа бифуркаций торов Лиувилля и построения инварианта Фоменко-Цишанга
\cite{bolfom1999}. Соответствующая бифуркационная диаграмма $\Sigma_1$ отображения
момента  ${\cal F}_1=H\times F_0$ изображена на рис.~1. Здесь же указаны бифуркации торов
Лиувилля.
\begin{figure}[!htbp]
\centering
\includegraphics[width=8cm,keepaspectratio]{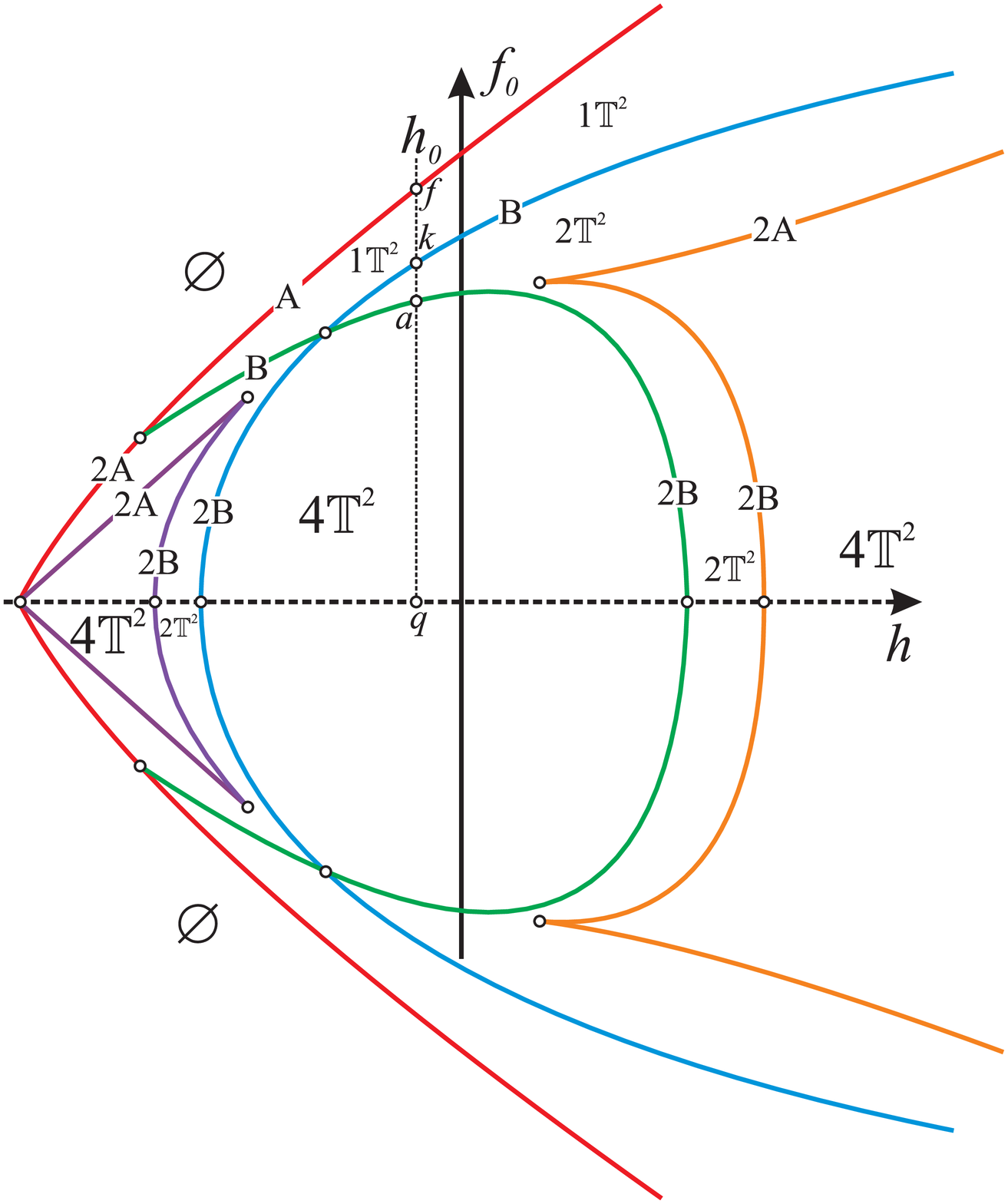}
\parbox[t]{0.9\textwidth}{\caption{Бифуркационная диаграмма $\Sigma_1$ отображения момента ${\cal F}_1=H\times F_0$.}\label{fig1}}
\end{figure}

При указанных значениях параметра деформации $\varepsilon$ параметризацию бифуркационной
диаграммы $\Sigma_1$ явно можно описать следующим образом:
\begin{equation*}\label{eq_2_10}
\delta_1:\left\{\begin{array}{l}
\ds{h=-\varepsilon^2(a^2+b^2)+2t(2+\varepsilon^2t)-\frac{2}{t}(1+\varepsilon^2t)\sqrt{(a^2-t^2)(b^2-t^2)},}\\[5mm]
\ds{f_0=\pm\frac{2}{t}(1+\varepsilon^2t)\sqrt{\varepsilon^2[\sqrt{(a^2-t^2)(b^2-t^2)}-t^2]-t}
\sqrt[4]{(a^2-t^2)(b^2-t^2)}(\sqrt{a^2-t^2}+\sqrt{b^2-t^2})},\\[5mm]
\ds{t\in[-\frac{1}{\varepsilon^2};0)},
\end{array}\right.
\end{equation*}
\begin{equation*}\label{eq_2_11}
\delta_2:\left\{\begin{array}{l}
\ds{h=-\varepsilon^2(a^2+b^2)+2t(2+\varepsilon^2t)+\frac{2}{t}(1+\varepsilon^2t)\sqrt{(a^2-t^2)(b^2-t^2)},}\\[5mm]
\ds{f_0=\pm\frac{2}{t}(1+\varepsilon^2t)\sqrt{\varepsilon^2[\sqrt{(a^2-t^2)(b^2-t^2)}+t^2]+t}
\sqrt[4]{(a^2-t^2)(b^2-t^2)}(\sqrt{a^2-t^2}-\sqrt{b^2-t^2})},\\[5mm]
\ds{t\in[-b;-\frac{1}{\varepsilon^2}]},
\end{array}\right.
\end{equation*}
\begin{equation*}\label{eq_2_12}
\delta_3:\left\{\begin{array}{l}
\ds{h=-\varepsilon^2(a^2+b^2)+2t(2+\varepsilon^2t)-\frac{2}{t}(1+\varepsilon^2t)\sqrt{(t^2-a^2)(t^2-b^2)},}\\[5mm]
\ds{f_0=\pm\frac{2}{t}(1+\varepsilon^2t)\sqrt{\varepsilon^2[t^2-\sqrt{(t^2-a^2)(t^2-b^2)}]+t}
\sqrt[4]{(t^2-a^2)(t^2-b^2)}(\sqrt{t^2-a^2}-\sqrt{t^2-b^2})},\\[5mm]
\ds{t\in[t_0;-a]}.
\end{array}\right.
\end{equation*}
Здесь через $t_0$ обозначено выражение
\begin{equation*}
t_0=-\frac{1}{8\varepsilon^2}[4+\varepsilon^4(a+b)^2+\sqrt{16+8(a^2+b^2-6ab)\varepsilon^4+(a+b)^4\varepsilon^8}].
\end{equation*}
\begin{equation*}\label{eq_2_13}
\delta_4:\left\{\begin{array}{l}
\ds{h=-\varepsilon^2(a^2+b^2)+2t(2+\varepsilon^2t)+\frac{2}{t}(1+\varepsilon^2t)\sqrt{(a^2-t^2)(b^2-t^2)},}\\[5mm]
\ds{f_0=\pm\frac{2}{t}(1+\varepsilon^2t)\sqrt{\varepsilon^2[\sqrt{(a^2-t^2)(b^2-t^2)}+t^2]+t}
\sqrt[4]{(a^2-t^2)(b^2-t^2)}(\sqrt{a^2-t^2}-\sqrt{b^2-t^2})},\\[5mm]
\ds{t\in(0;b]},
\end{array}\right.
\end{equation*}
\begin{equation*}\label{eq_2_14}
\delta_5:\left\{\begin{array}{l}
\ds{h=-\varepsilon^2(a^2+b^2)+2t(2+\varepsilon^2t)-\frac{2}{t}(1+\varepsilon^2t)\sqrt{(t^2-a^2)(t^2-b^2)},}\\[5mm]
\ds{f_0=\pm\frac{2}{t}(1+\varepsilon^2t)\sqrt{\varepsilon^2[t^2-\sqrt{(t^2-a^2)(t^2-b^2)}]+t}
\sqrt[4]{(t^2-a^2)(t^2-b^2)}(\sqrt{t^2-b^2}-\sqrt{t^2-a^2})},\\[5mm]
\ds{t\in[a,+\infty)}.
\end{array}\right.
\end{equation*}

\section{Вторая система}

По-прежнему, предполагаем $\lambda=0$. Рассмотрим функциональное равенство, которое
выполняется тождественно в точках фазового пространства $\mP^6$:
\begin{equation}\label{eq_3_1}
[2+\varepsilon^4({\boldsymbol\alpha}^2+{\boldsymbol\beta}^2)+\varepsilon^2H]^2-4\varepsilon^4K=Q_1^2+Q_2^2+Q_3^2,
\end{equation}
где
\begin{equation*}\label{eq_3_2}
\begin{array}{l}
\ds{Q_1=2\varepsilon[(\alpha_3\beta_2-\alpha_2\beta_3)\varepsilon^3+(\alpha_2M_1+\beta_2M_2+\beta_3M_3)\varepsilon^2-
(M_1M_3+\alpha_3)\varepsilon-M_2],}\\[5mm]
\ds{Q_2=2\varepsilon[(\alpha_3\beta_1-\alpha_1\beta_3)\varepsilon^3+(\alpha_1M_1+\beta_1M_2+\alpha_3M_3)\varepsilon^2+
(M_2M_3+\beta_3)\varepsilon-M_1],}\\[5mm]
\ds{Q_3=2[(\alpha_1\beta_2-\alpha_2\beta_1)\varepsilon^4+\varepsilon^3(\beta_1-\alpha_2)M_3+
(M_3^2-\alpha_1-\beta_2)\varepsilon^2+1].}
\end{array}
\end{equation*}

Выберем постоянные первых интегралов, удовлетворяющих соотношению
\begin{equation*}\label{eq_3_3}
k=\frac{1}{4\varepsilon^4}[2+\varepsilon^4(a^2+b^2)+\varepsilon^2h]^2.
\end{equation*}
Тогда соотношение (\ref{eq_3_1}) приводит к системе уравнений
\begin{equation}\label{eq_3_4}
Q_k=0,\quad k=1,2,3,
\end{equation}
которая равносильна одной из систем:

либо
\begin{equation}\label{eq_3_5}
Q_1=0,\quad Q_3=0,
\end{equation}

либо
\begin{equation}\label{eq_3_6}
Q_2=0,\quad Q_3=0.
\end{equation}
Ограничимся выбором параметра деформации
$\varepsilon>\max\{\frac{1}{\sqrt{b}};\sqrt{\frac{2}{a-b}}\}$. В этом случае системам
(\ref{eq_3_4}) -- (\ref{eq_3_6}) удовлетворяют положения равновесия $c_1$--$c_8$ и
$c_{13}$ -- $c_{16}$. Исключим из рассмотрения указанные положения равновесия. Тогда
имеет место
\begin{propos}
Система соотношений (\ref{eq_3_5}) определяет инвариантное четырехмерное подмногообразие
$\mn$ фазового пространства $\mP^6$ уравнений (\ref{eq_1_0}) с гамильтонианом
(\ref{eq_2_4}).
\end{propos}
\begin{proof}
Действительно,  справедливы равенства
\begin{equation*}\label{eq_3_7}
\begin{array}{l}
\ds{\dot Q_1=\{H,Q_1\}=a_{11}Q_1+a_{12}Q_3,}\\[5mm]
\ds{\dot Q_3=\{H,Q_3\}=a_{21}Q_1+a_{22}Q_3,}\\[5mm]
\end{array}
\end{equation*}
где
\begin{equation*}\label{eq_3_8}
\begin{array}{l}
\ds{a_{11}=\frac{2}{\varepsilon(\alpha_2\varepsilon-M_3)}[\alpha_1(\alpha_2-\beta_1)\varepsilon^3-
\varepsilon^2M_3\alpha_1+\varepsilon(\beta_1-\alpha_2)+M_3],}\\[5mm]
\ds{a_{12}=\frac{2}{\alpha_2\varepsilon-M_3}[\varepsilon^2\alpha_3\beta_1+\varepsilon(\beta_1-\alpha_2)M_2
+M_2M_3],}\\[5mm]
\ds{a_{21}=-\frac{2}{\alpha_2\varepsilon-M_3}[(\alpha_2\alpha_3-\alpha_1\beta_3)\varepsilon^2-
\varepsilon\alpha_3M_3+\beta_3],}\\[5mm]
\ds{a_{22}=-\frac{2\varepsilon\beta_3(M_2+\varepsilon\alpha_3)}{\alpha_2\varepsilon-M_3}.}
\end{array}
\end{equation*}
В силу (\ref{eq_3_5}) имеем: ${\dot Q_1}=0$, ${\dot Q_3}=0$.
\end{proof}

В точках подмногообразия $\mn$ фазового пространства $\mP^6$ выполняются соотношения
\begin{equation}\label{eq_3_9}
\begin{array}{l}
\ds{\varepsilon^2[\varepsilon^4(G+2\varepsilon^2{\boldsymbol\alpha}^2{\boldsymbol\beta}^2)-H]-2=\varepsilon^4F_1^2,}\\[5mm]
\ds{\dot F_1=\{H,F_1\}=0,}
\end{array}
\end{equation}
 где выражение для $F_1$ определено формулой

\begin{equation*}\label{eq_3_10}
F_1=\frac {\sqrt{(\beta_2\varepsilon^2-1)^2+\varepsilon^2(M_3-\varepsilon\alpha_2)^2}
\cdot A}{ \left( {\varepsilon}^{2} \beta_{{2}}-1 \right) \left(
\alpha_{{2}}\varepsilon-M_{{3}} \right)
 \left( \alpha_{{1}}-\beta_{{2}} \right) {\varepsilon}^{2}},
\end{equation*}
\begin{equation*}
\begin{array}{l}
\ds{A=\left(
{\alpha_{{2}}}^{2}\alpha_{{1}}\beta_{{3}}+\alpha_{{3}}{\beta_{{2}}}^{2}\alpha_{{2}}-\alpha_{{3}}\alpha_{{2}}\alpha_{{1}}\beta_{{2}}+
\alpha_{{3}}{\beta_{{3}}}^{2}\alpha_{{2}}+\beta_{{3}}\beta_{{2}}{ \alpha_{{1}}}^{2}
\right) {\varepsilon}^{5}+}\\[5mm]
\ds{+\left( -2\,\alpha_{{2}}M_{{3
}}\alpha_{{1}}\beta_{{3}}-\alpha_{{3}}\beta_{{3}}\alpha_{{2}}M_{{1}}-
\alpha_{{3}}{\beta_{{3}}}^{2}M_{{3}}+\alpha_{{3}}M_{{3}}\alpha_{{1}}
\beta_{{2}}-\alpha_{{3}}{\beta_{{2}}}^{2}M_{{3}}-{\beta_{{2}}}^{2}
\alpha_{{1}}M_{{1}}-{\alpha_{{2}}}^{2}\alpha_{{1}}M_{{1}} \right) {
\varepsilon}^{4}+}\\[5mm]
\ds{+\left( \alpha_{{3}}\beta_{{3}}M_{{3}}M_{{1}}+\alpha_{{3
}}\alpha_{{2}}\alpha_{{1}}+{M_{{3}}}^{2}\alpha_{{1}}\beta_{{3}}-\beta_
{{3}}{\alpha_{{1}}}^{2}+2\,\alpha_{{2}}M_{{3}}\alpha_{{1}}M_{{1}}-
\alpha_{{1}}\beta_{{2}}\beta_{{3}}-\alpha_{{3}}\beta_{{2}}\alpha_{{2}}
 \right) {\varepsilon}^{3}+}\\[5mm]
\ds{+\left( -\alpha_{{3}}M_{{3}}\alpha_{{1}}-{M_{{
3}}}^{2}\alpha_{{1}}M_{{1}}+\alpha_{{3}}\beta_{{2}}M_{{3}}+{\beta_{{2}
}}^{2}M_{{1}}+\alpha_{{1}}\beta_{{2}}M_{{1}} \right)
{\varepsilon}^{2}+\alpha_{{1}}\varepsilon\,\beta_{{3}}-\beta_{{2}}M_{{1}}.}
\end{array}
\end{equation*}

Заметим, что $a_{11}+a_{22}\ne 0$, поэтому скобка $\{Q_1,Q_3\}$ не является интегралом. В
качестве второго дополнительного интеграла можно взять функцию $G$ из (\ref{eq_2_1}) или,
в силу (\ref{eq_3_9}), функцию $F_1$. На рис.~2 изображены бифуркационная диаграмма
$\Sigma_2$ отображения момента  ${\cal F}_2=H\times G$ и бифуркации торов Лиувилля.

\begin{figure}[!htbp]
\centering
\includegraphics[width=12cm,keepaspectratio]{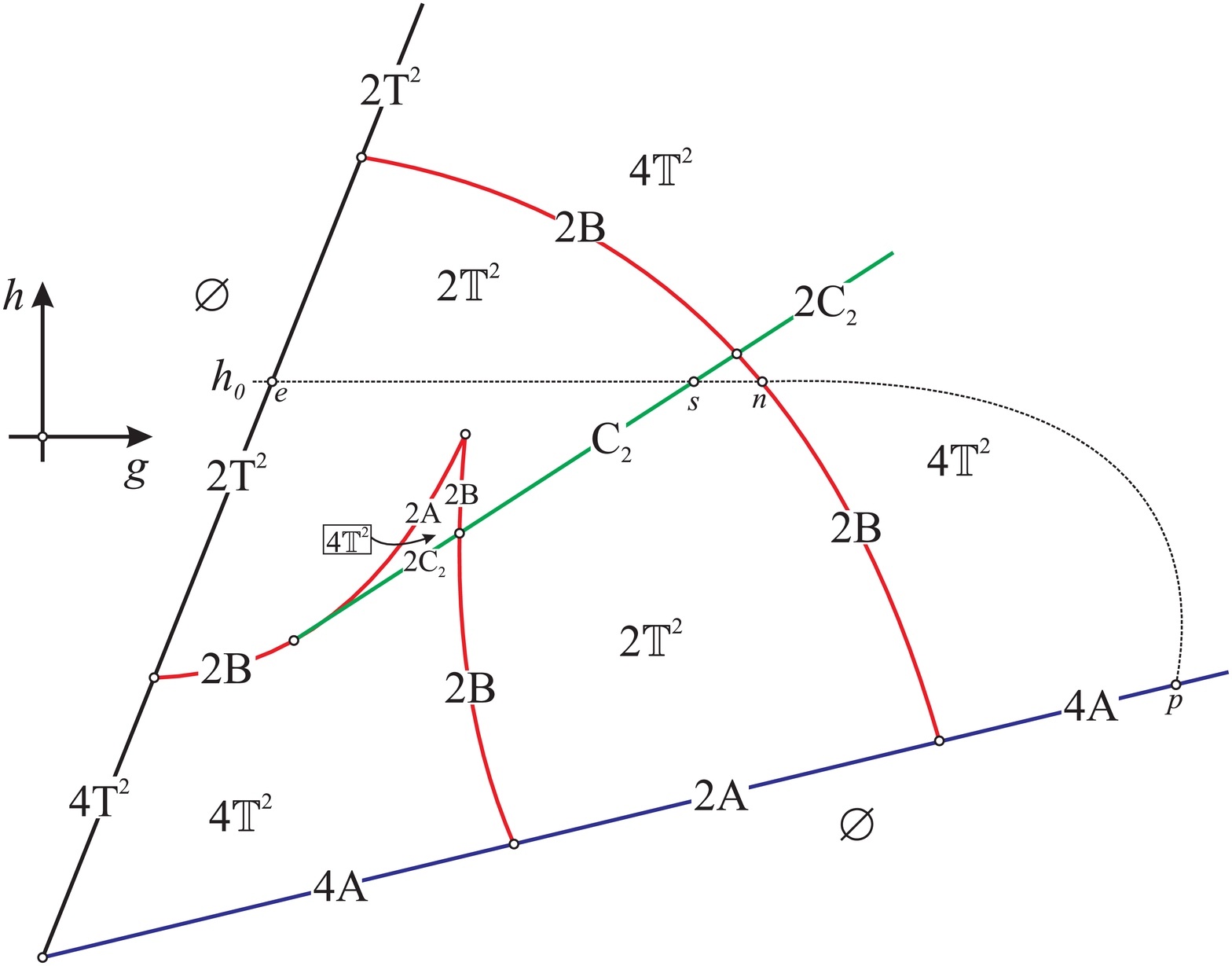}
\parbox[t]{0.9\textwidth}{\caption{Бифуркационная диаграмма $\Sigma_2$ отображения момента ${\cal F}_2=H\times G$.}\label{fig2}}
\end{figure}

Параметризация бифуркационной диаграммы $\Sigma_2$ описывается следующими кривыми:
\begin{equation*}\label{eq_3_11}
\Sigma_2:\quad\begin{array}{l}
\ds{h=\frac{g}{b^2}+(\varepsilon^2+\frac{1}{\varepsilon^2b^2})(a^2-b^2),\quad
g\geqslant-\varepsilon^2b^2(a^2+b^2)-\frac{2a^2}{\varepsilon^2},}\\[5mm]
\ds{g=\frac{h}{\varepsilon^4}+\frac{2}{\varepsilon^6}-2\varepsilon^2a^2b^2,\quad
h\geqslant-\frac{1}{\varepsilon^2}[\varepsilon^4(a^2+b^2)+2],}\\[5mm]
\ds{h=\frac{g}{a^2}-(\varepsilon^2+\frac{1}{\varepsilon^2a^2})(a^2-b^2),\quad g\geqslant
-\frac{1}{\varepsilon^2}(a^2+b^2+2\varepsilon^4a^2b^2),}\\[5mm]
\left\{\begin{array}{l}
\ds{h=2t-\frac{1}{\varepsilon^2}-\frac{\varepsilon^2a^2b^2}{t^2},}\\[5mm]
\ds{g=-\varepsilon^2a^2b^2-\frac{t^2}{\varepsilon^2}+\frac{2a^2b^2}{t}, \quad
t\in[-\varepsilon^2ab;-b]\cup[b;\varepsilon^2ab].}
\end{array}\right.
\end{array}
\end{equation*}

\section{Третья и четвертая системы}
Заметим, что первые две подсистемы $\mm$ и  $\mn$ удовлетворяют соотношениям
\begin{equation}\label{eq_4_0}
L_i(K,H,G)=0,\quad dL_i(K,H,G)=0,\quad i=1,2,
\end{equation}
если положить $L_1=K$ и
$L_2=[2+\varepsilon^4({\boldsymbol\alpha}^2+{\boldsymbol\beta}^2)+\varepsilon^2H]^2-4\varepsilon^4K$
соответственно.

Поэтому последние две подсистемы  $\mo$ и $\ml$ определим уравнениями:
\begin{equation}\label{eq_4_1}
L_i(K,H,G)=0,\quad dL_i(K,H,G)=0,\quad i=3,4.
\end{equation}
Систему $\mo$ определим выбором функции
\begin{equation*}\label{eq_4_2}
L_3=[(a^2+b^2)H-2G-\varepsilon^2(a^2-b^2)^2]^2-4(a^2-b^2)^2K,
\end{equation*}
а функцию $L_4$ для $\ml$ можно явно записать, исключив из системы
\begin{equation}\label{eq_4_3}
\Phi(t)=0,\quad \frac{d\Phi(t)}{dt}=0
\end{equation}
параметр $t$ и подставив в полученное выражение вместо постоянных первых интегралов
$h,k,g$ функции $H,K$ и $G$. Здесь через $\Phi(t)$ обозначен многочлен
\begin{equation*}\label{eq_4_4}
\Phi(t)=t^4-ht^3+\left(\frac{1}{4}[h^2+2\varepsilon^2(a^2+b^2)h+\varepsilon^4(a^2-b^2)^2]
+a^2+b^2-k-\varepsilon^2g\right)t-gt+a^2b^2.
\end{equation*}
Параметр $t$ в результате исключения из системы (\ref{eq_4_3}) как функция $T=T(H,K,G)$,
на самом деле, является частным интегралом для системы $\ml$.

Отметим, что уравнению $L_4(h,k,g)=0$ принадлежат две прямые
\begin{equation*}\label{eq_4_5}
\left\{\begin{array}{l} \ds{g=abh,}\\[5mm]
\ds{k=\frac{1}{4}(a-b)^2[\varepsilon^4(a+b)^2+2\varepsilon^2h+4],} \end{array}\right.
\left\{\begin{array}{l} \ds{g=-abh,}\\[5mm]
\ds{k=\frac{1}{4}(a+b)^2[\varepsilon^4(a-b)^2+2\varepsilon^2h+4].} \end{array}\right.
\end{equation*}

Систему (\ref{eq_4_3}) (поверхность кратных корней многочлена $\Phi(t)$) можно записать в
параметрическом виде следующим образом:
\begin{equation*}\label{eq_4_6}
\left\{\begin{array}{l} \ds{k=3t^2-2th+a^2+b^2+\frac{1}{4}h^2-\frac{a^2b^2}{t^2}+}\\[5mm]
\ds{+[2t^3-ht^2+\frac{1}{2}(a^2+b^2)h-\frac{2a^2b^2}{t}]\varepsilon^2+\frac{1}{4}(a^2-b^2)^4\varepsilon^4,}\\[5mm]
\ds{g=-2t^3+ht^2+\frac{2a^2b^2}{t},\quad t\in{\Bbb R}\setminus 0.}
\end{array}\right.
\end{equation*}

Если в приводимых выше формулах положить параметр деформации $\varepsilon$ равным нулю,
то получим формулы для аналогов систем $\mo$ и $\ml$ в задаче обобщения волчка
Ковалевской на случай двойного поля \cite{Kh34}, \cite{KhSav}, \cite{Kh2005},
\cite{Kh2006}, \cite{Kh2007}, \cite{Kh2009}, \cite{Kh2011}.

При $\varepsilon>\max\{\frac{1}{\sqrt{b}};\sqrt{\frac{2}{a-b}}\}$ бифуркационные
диаграммы $\Sigma_3$, $\Sigma_4$ отображений моментов ${\cal F}_3=G\times H$  и ${\cal
F}_4=T\times H$ для систем $\mo$ и $\ml$ изображены на рис.~3 и 4 соответственно. На этих
же рисунках указаны и бифуркации торов Лиувилля. В основе анализа бифуркациий торов
Лиувилля лежат аналитические работы \cite{Kh34}, \cite{KhSav}, \cite{Kh2005},
\cite{Kh2006}, \cite{Kh2007}, \cite{Kh2009}, \cite{Kh2011} и \cite{KhRyab_ms2012}.
Закрашенным областям соответствуют устойчивые двухчастотные периодические решения или, в
другой терминологии, невырожденные особенности ранга 2 эллиптического типа полного
отображения момента ${\cal F}=H\times K\times G$, а остальная часть --- двухчастотным
периодическим решениям (двумерным торам) гиперболического типа по отношению ко всей
системе с тремя степенями свободы. Построения велись для следующих значений параметров:
$a=1,b=\frac{2}{5},\varepsilon=3$.

\begin{figure}[!htbp]
\centering
\includegraphics[width=13cm,keepaspectratio]{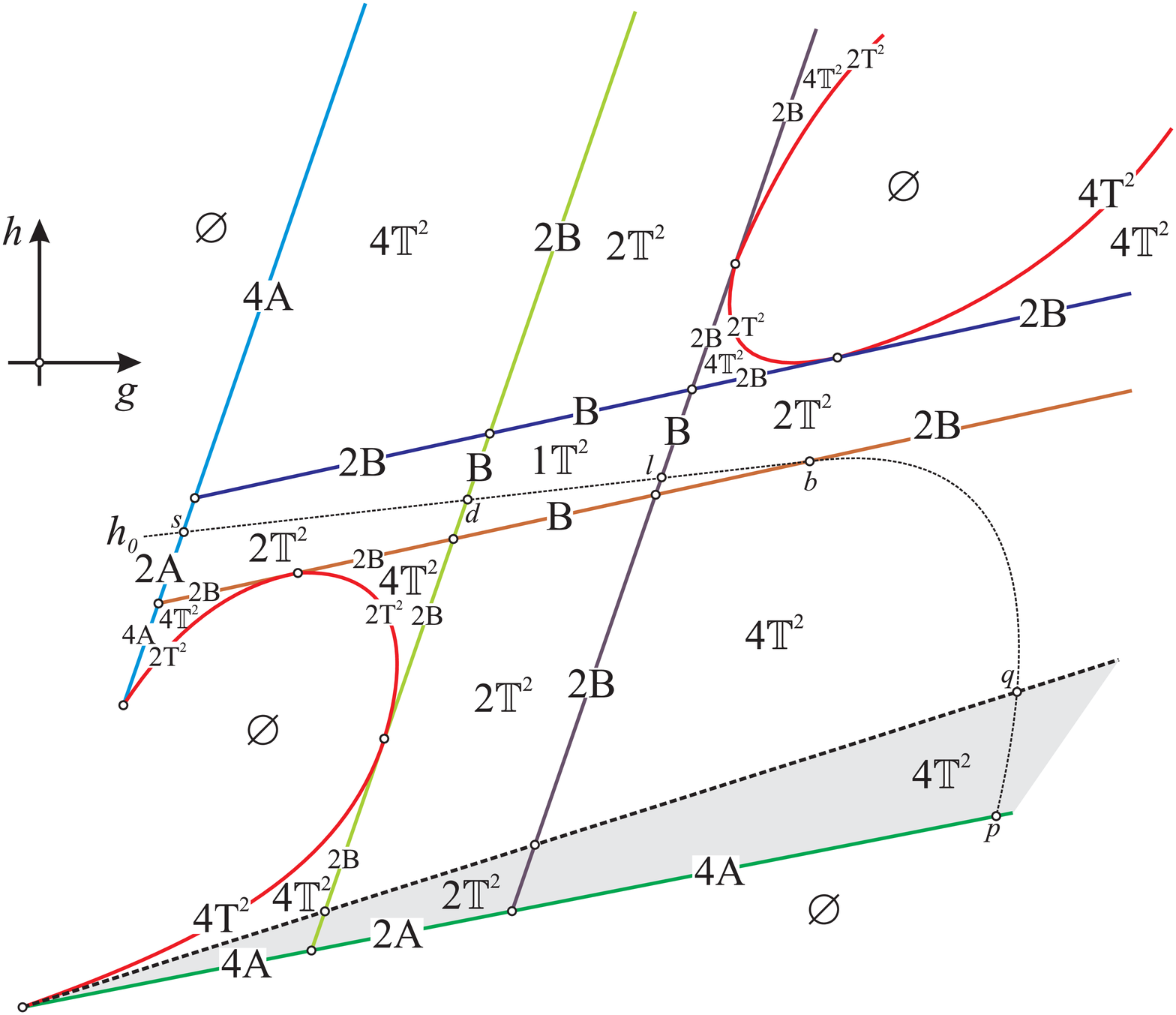}
\parbox[t]{0.9\textwidth}{\caption{Бифуркационная диаграмма $\Sigma_3$ отображения момента ${\cal F}_3=G\times H$.}\label{fig3}}
\end{figure}

\begin{figure}[!htbp]
\centering
\includegraphics[width=14cm,keepaspectratio]{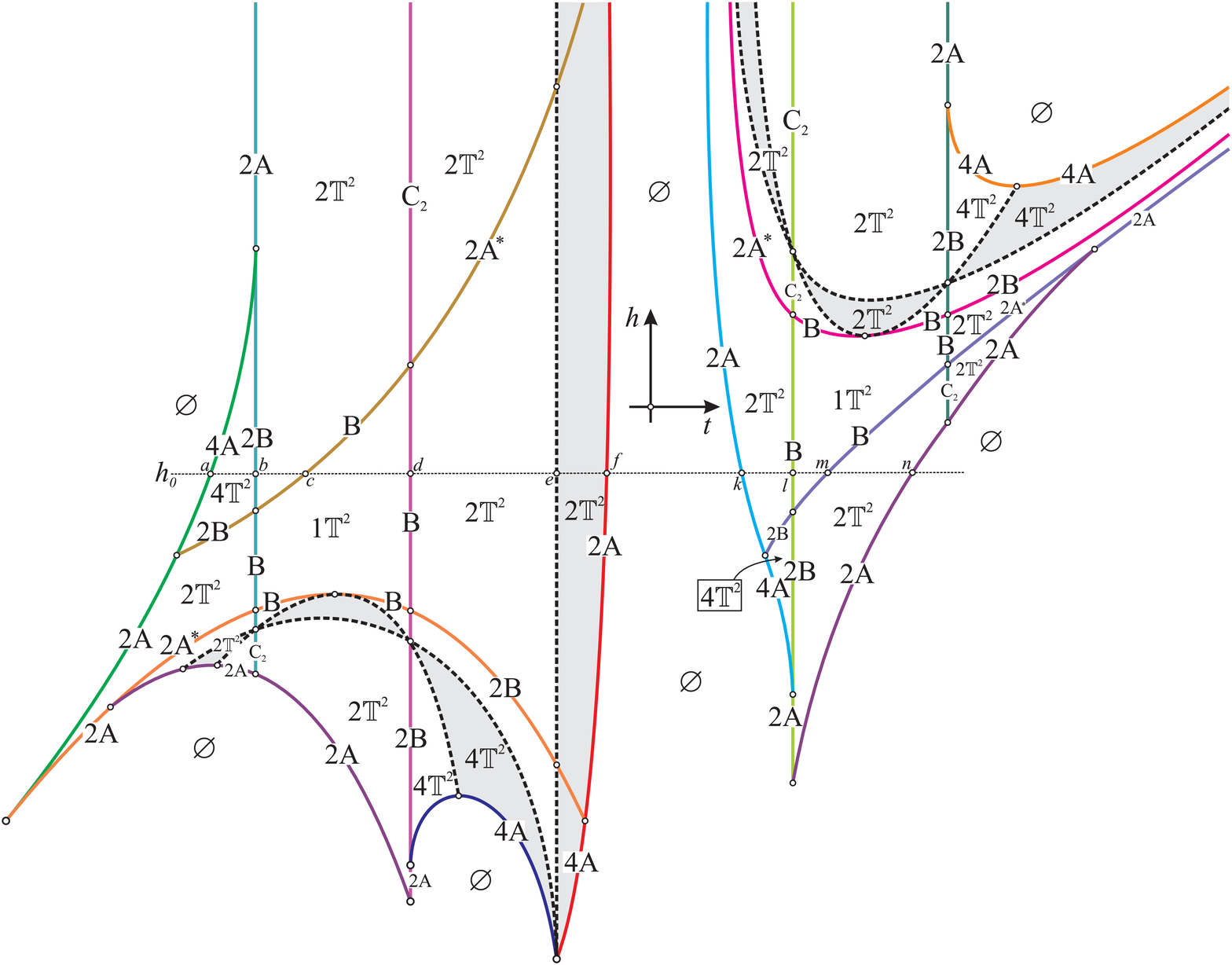}
\parbox[t]{0.9\textwidth}{\caption{Бифуркационная диаграмма $\Sigma_4$ отображения момента ${\cal F}_4=T\times H$.}\label{fig4}}
\end{figure}

Предъявим явное описание (параметризацию) бифуркационной диаграммы $\Sigma_3$:

\begin{equation*}\label{eq_4_7}
\Sigma_3:\quad\begin{array}{l}
\ds{h=\frac{g}{a^2}-(\varepsilon^2+\frac{1}{\varepsilon^2a^2})(a^2-b^2),\quad g\geqslant
-\frac{1}{\varepsilon^2}(a^2+b^2+2\varepsilon^4a^2b^2),}\\[5mm]
\left\{\begin{array}{l} \ds{h=2t+\frac{a^2+b^2}{t},}\\[5mm]
\ds{g=(a^2+b^2)t+\frac{2a^2b^2}{t},\quad
t\in[-\varepsilon^2b^2;-\frac{1}{\varepsilon^2}]\cup (0,;+\infty),}
\end{array}\right.\\[10mm]
\ds{g=a^2h+2a(a^2-b^2),\quad
h\geqslant-\frac{1+2\varepsilon^2a+\varepsilon^4b^2}{\varepsilon^2},}\\[5mm]
\ds{g=a^2h-2a(a^2-b^2),\quad
h\geqslant-\frac{1-2\varepsilon^2a+\varepsilon^4b^2}{\varepsilon^2},}\\[5mm]
\ds{h=\frac{g}{b^2}+(\varepsilon^2+\frac{1}{\varepsilon^2b^2})(a^2-b^2),\quad
g\geqslant-\varepsilon^2b^2(a^2+b^2)-\frac{2a^2}{\varepsilon^2},}\\[5mm]
\ds{g=b^2h-2b(a^2-b^2),\quad
h\geqslant-\frac{1+2\varepsilon^2b+\varepsilon^4a^2}{\varepsilon^2},}\\[5mm]
\ds{g=b^2h+2b(a^2-b^2),\quad
h\geqslant-\frac{1-2\varepsilon^2b+\varepsilon^4a^2}{\varepsilon^2}.}
\end{array}
\end{equation*}

Пересечение систем $\mm$ и $\mo$ происходит в точках вырождения индуцированной
симплектической структуры, что соответствует на рис.~3 прямой
\begin{equation*}\label{eq_4_8}
h=\frac{1}{a^2+b^2}[2g-\varepsilon^2(a^2-b^2)^2],\quad
g\geqslant-\frac{1}{\varepsilon^2}(a^2+b^2+2\varepsilon^4a^2b^2).
\end{equation*}

Параметризация бифуркационной диаграммы  $\Sigma_4$ явно описывается системой кривых:
\begin{equation}\label{eq_4_9}
\Sigma_4:\quad\begin{array}{l}
\ds{h=-\varepsilon^2(a^2+b^2)+2t(2+\varepsilon^2t)-\frac{2}{t}(1+\varepsilon^2t)\sqrt{(a^2-t^2)(b^2-t^2)},\quad t\in[-\frac{1}{\varepsilon^2};0),}\\[5mm]
\ds{h=-\varepsilon^2(a^2+b^2)+2t(2+\varepsilon^2t)+\frac{2}{t}(1+\varepsilon^2t)\sqrt{(a^2-t^2)(b^2-t^2)},\quad t\in[-b;-\frac{1}{\varepsilon^2}]\cup (0;b],}\\[5mm]
\ds{h=-\varepsilon^2(a^2+b^2)+2t(2+\varepsilon^2t)-\frac{2}{t}(1+\varepsilon^2t)\sqrt{(t^2-a^2)(t^2-b^2)},\quad t\in[t_0;-a]\cup [a;+\infty),}\\[5mm]
\ds{h=2t+\frac{2ab}{t},\quad t\in[t_0;t_1]\cup (0;+\infty);\quad
h=2t-\frac{2ab}{t},\quad t\in[t_3;0)\cup [t_2;+\infty),}\\[5mm]
\ds{h=2t-\frac{\varepsilon^2a^2b^2}{t^2}-\frac{1}{\varepsilon^2},\quad t\in[-\varepsilon^2ab;-b]\cup[b;\varepsilon^2ab],}\\[5mm]
\ds{t=\pm a,\quad h\geqslant-\frac{1\mp
2\varepsilon^2a+\varepsilon^4b^2}{\varepsilon^2};\quad
t=\pm b,\quad h\geqslant-\frac{1\mp 2\varepsilon^2b+\varepsilon^4a^2}{\varepsilon^2}.}\\[5mm]
\end{array}
\end{equation}

В формулах (\ref{eq_4_9}) через $t_k$ обозначены выражения
\begin{equation*}\label{eq_4_10}
\begin{array}{l}
\ds{t_{0,1}=-\frac{1}{8\varepsilon^2}[\varepsilon^4(a+b)^2+4\pm\sqrt{16+\varepsilon^8(a+b)^4+8(a^2+b^2-6ab)\varepsilon^4}],}\\[5mm]
\ds{t_{2,3}=-\frac{1}{8\varepsilon^2}[\varepsilon^4(a-b)^2+4\mp\sqrt{16+\varepsilon^8(a-b)^4+8(a^2+b^2+6ab)\varepsilon^4}].}\\[5mm]

\end{array}
\end{equation*}

Закрашенная область ограничена кривыми
\begin{equation*}\label{eq_4_11}
\begin{array}{l}
\ds{t=-\frac{1}{\varepsilon^2},\quad h\geqslant h_{\min}=-\frac{\varepsilon^4(a^2+b^2)+2}{\varepsilon^2},}\\[5mm]
\ds{h=2t+\frac{a^2+b^2}{t},\quad t\in[-\varepsilon^2b^2;-\frac{1}{\varepsilon^2}]\cup
(0;+\infty),}\\[5mm]
\ds{h=3t+\frac{a^2b^2}{t^3},\quad t\in[-\sqrt[3]{\varepsilon^2a^2b^2};t_4]\cup (0;t_5]},
\end{array}
\end{equation*}
где $t_{4,5}$ -- единственные корни уравнений
\begin{equation*}\label{eq_4_12}
\begin{array}{l}
\ds{t^3(a^2+b^2-2t^2)\varepsilon^2-t^4+a^2b^2=2t^2(1+\varepsilon^2t)\sqrt{(a^2-t^2)(b^2-t^2)},\quad
t\in[-b;0),}\\[5mm]
\ds{t^4
-t^3(a^2+b^2-2t^2)\varepsilon^2-a^2b^2=2t^2(1+\varepsilon^2t)\sqrt{(t^2-a^2)(t^2-b^2)},\quad
t\in[a;+\infty).}\\[5mm]
\end{array}
\end{equation*}

\section{Атлас бифуркационных диаграмм и пример сетевой диаграммы}

В этом разделе приводится атлас бифуркационных диаграмм полного отображения момента
${\cal F}=H\times K\times G$  и его фрагмент. На рис.~5. по оси абсцисс откладывается
уровень энергии $h$, по оси ординат -- отношение $\frac{b}{a}$. Атлас построен для
значений параметров: $a=1, \varepsilon=2$. Все кривые, которые формируют атлас,
аналитически определяются по бифуркационным диаграммам $\Sigma_k, k=1,...,4$.

\begin{figure}[!htbp]
\centering
\includegraphics[width=6cm,keepaspectratio]{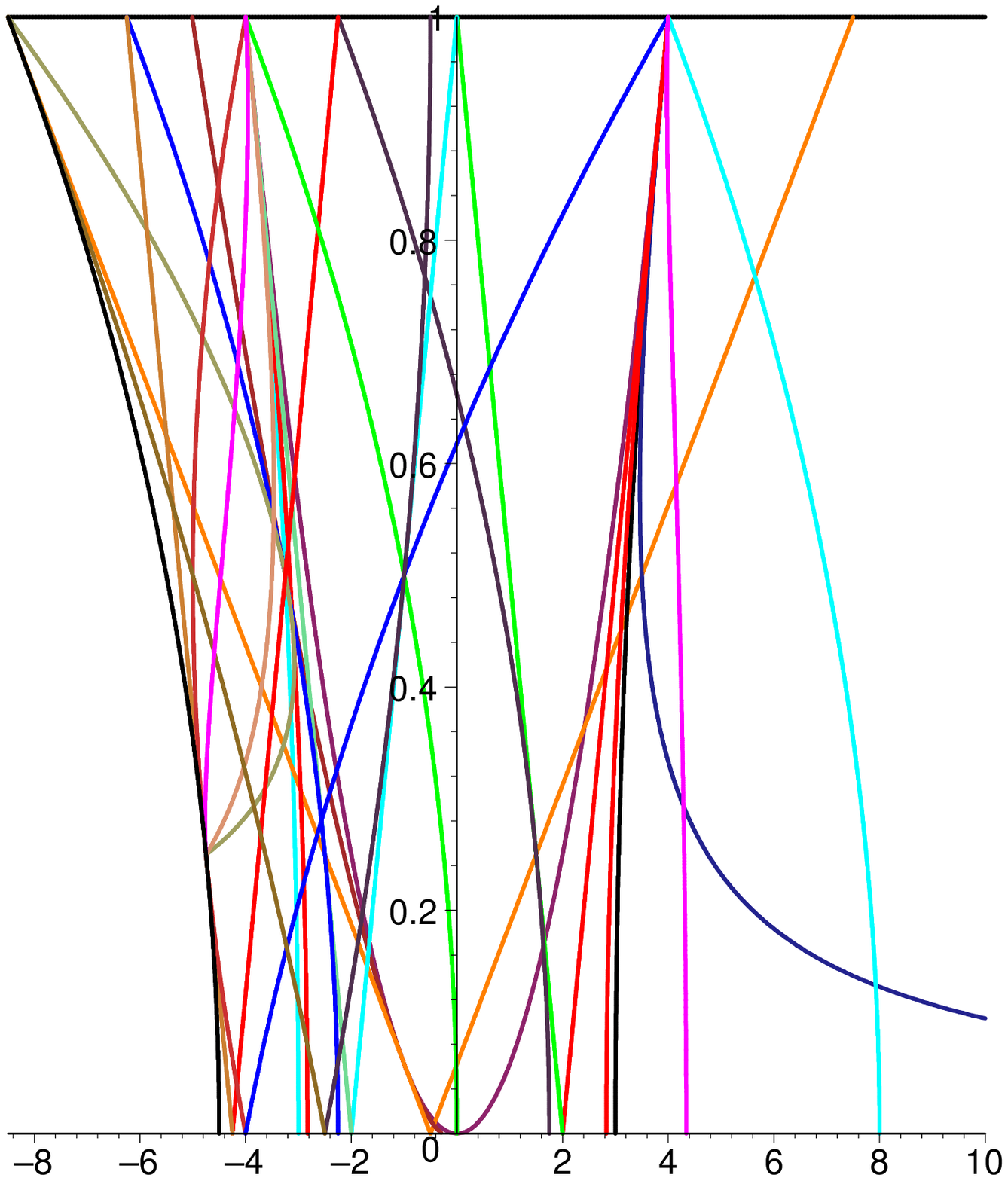}\quad
\includegraphics[width=6cm,keepaspectratio]{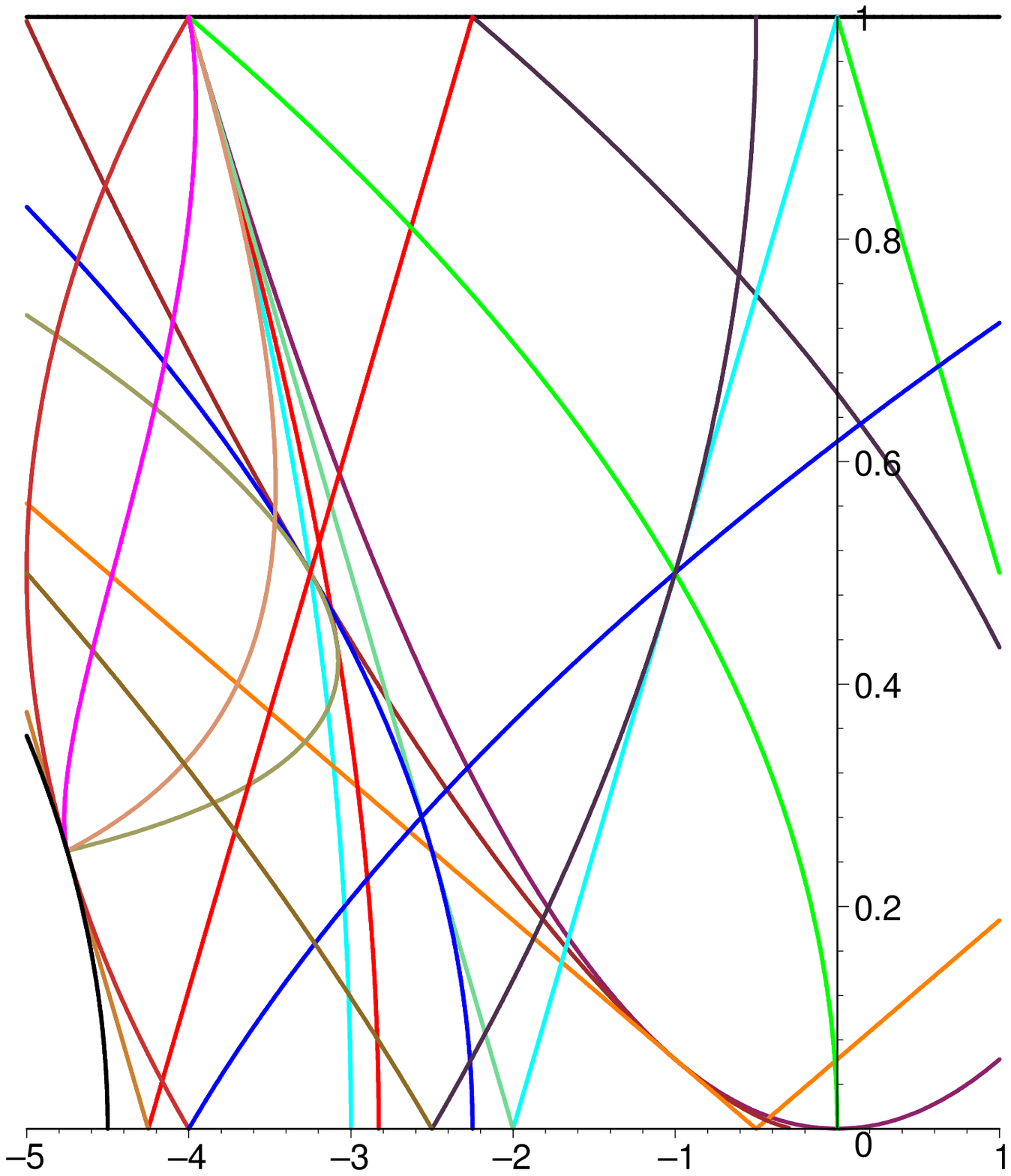}
\parbox[t]{0.9\textwidth}{\caption{Атлас бифуркационных диаграмм полного отображения момента \mbox{${\cal F}=H\times K\times G$}  и его фрагмент.}\label{fig5}}
\end{figure}

Для построения бифуркационной диаграммы $\Sigma_{h_0}$ множества $\Sigma$ полного
отображения момента ${\cal F}=H\times K\times G$ на уровне $h=h_0$ необходимо
зафиксировать значения параметров $a,\varepsilon$. Тем самым определено сечение атласа с
камерами. Для определенной камеры выбираем значения параметров $b$ и $h_0$ (сечение
изоэнергетической поверхности).  На бифуркационных диаграммах $\Sigma_k, k=1,\dots,4$
(рис.~1 -- 4) это сечение отмечено пунктирной линией. На указанных диаграммах для
выбранного сечения $h_0$ аналитически вычисляются границы, по которым и строится
бифуркационная диаграмма $\Sigma_{h_0}$.

На рис.~6 бифуркационная диаграмма $\Sigma_{h_0}$ построена для следующих значений
параметров: $a=1$, $b=\frac{2}{5}$, $\varepsilon=3,h_0=-0,25$. На стенках камер указаны
как количество компонент (двумерных торов), так и соответствующие бифуркации в системе с
тремя степенями свободы. Для промежутка $(f,q)$, который отвечает системе $\mm$,
количество двумерных торов необходимо удвоить. Пример сетевой диаграммы для выбранного
изоэнергетического сечения (аналога многомерной сети Фоменко \cite{fom91},
\cite{fomams91}) указан на рис.~7. Такой сетевой диаграммы нет среди сетевых диаграмм в
задаче о движении волчка Ковалевской в двойном поле, полный список которых представлен в
\cite{khryab2012}. Поэтому никакими заменами переменных рассматриваемую систему нельзя
преобразовать в известные.


\begin{figure}[!htbp]
\centering
\includegraphics[width=17cm,keepaspectratio]{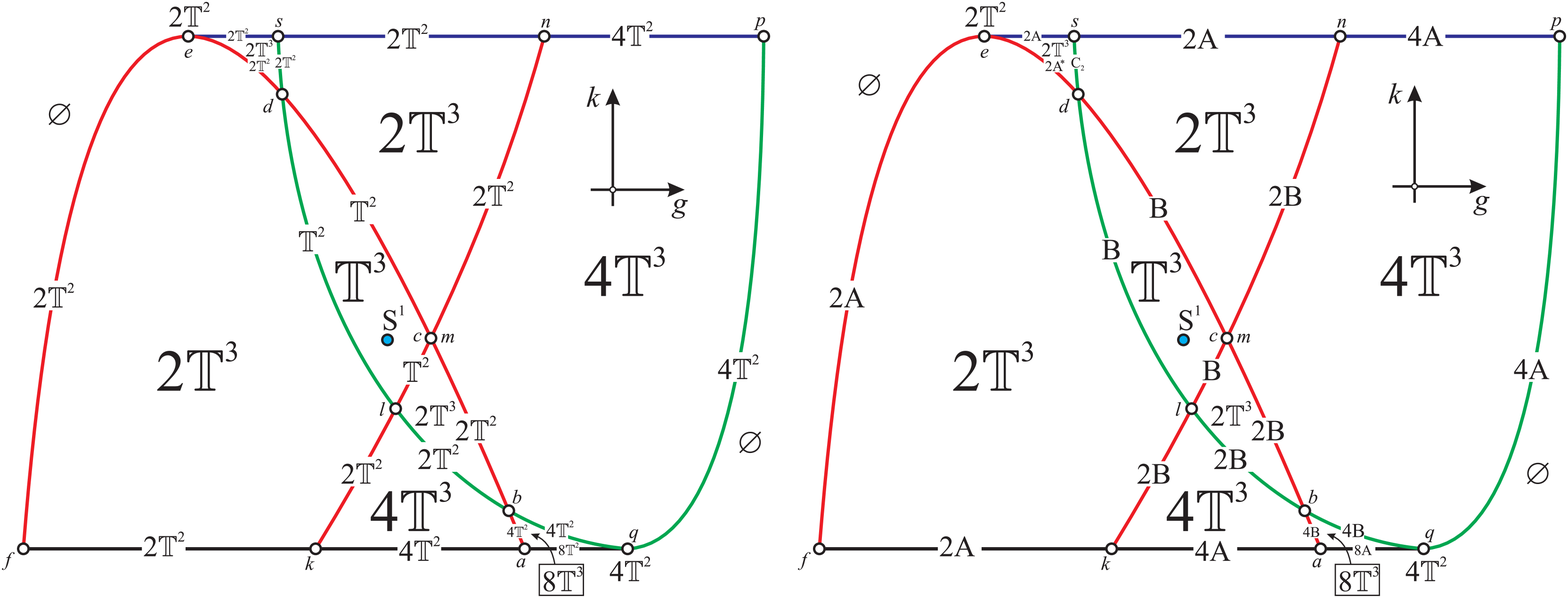}\qquad
\parbox[t]{0.9\textwidth}{\caption{Бифуркационная диаграмма $\Sigma_{h_0}$ для значений параметров \mbox{$a=1,
b=\frac{2}{5},\varepsilon=3,h_0=-0,25$.}}\label{fig6}}
\end{figure}

\begin{figure}[!htbp]
\centering
\includegraphics[width=9cm,keepaspectratio]{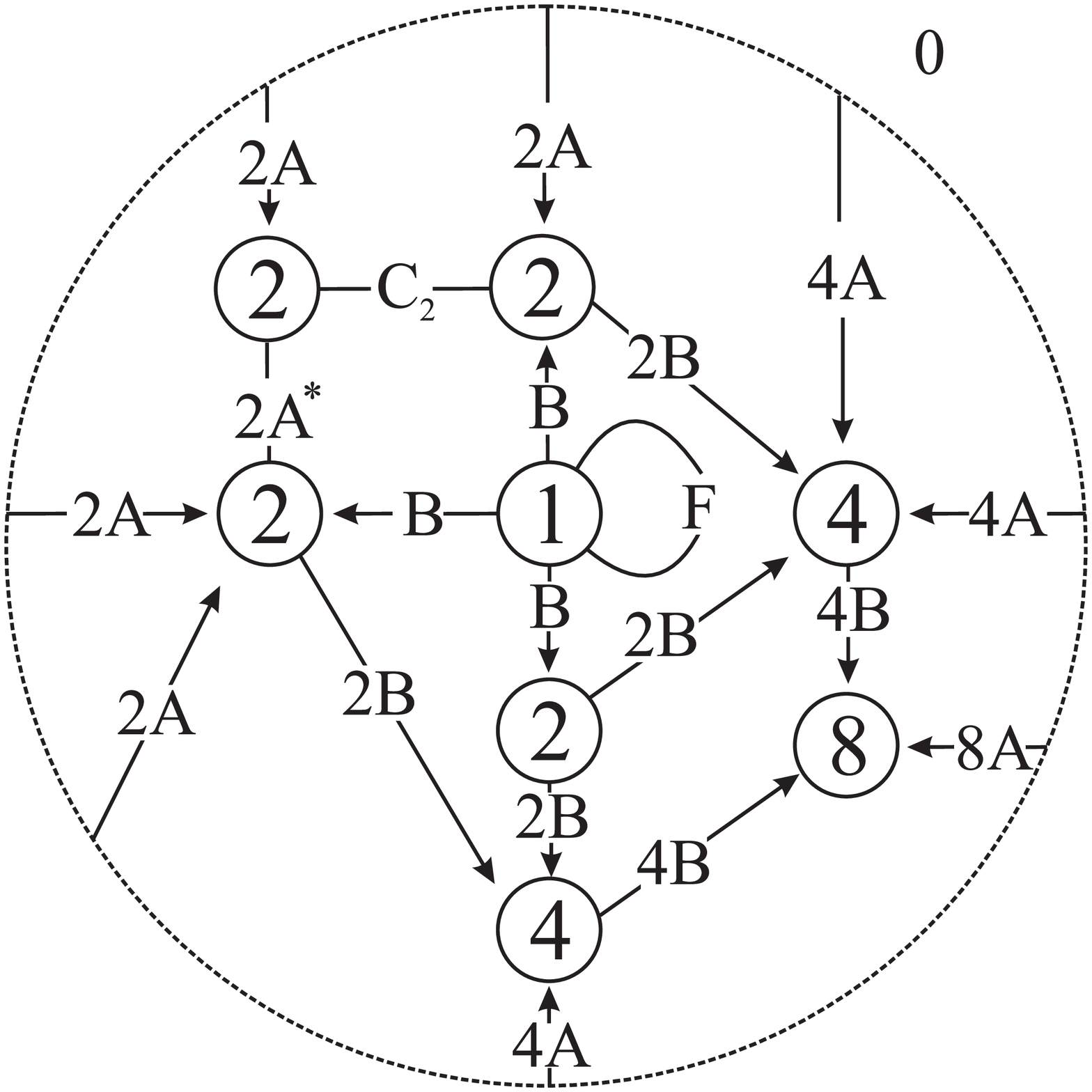}\qquad
\parbox[t]{0.9\textwidth}{\caption{Сетевая диаграмма.}\label{fig7}}
\end{figure}


\section{Заключение}

 В данной статье предложен возможный подход к описанию
фазовой топологии новой интегрируемой системы с тремя степенями свободы, используя метод
критических подсистем.  Понятие критической подсистемы введено М.\,П.~Харламовым в начале
2000-х годов в задаче исследования фазовой топологии неприводимых систем с тремя
степенями свободы. Основу результатов, анонсированных в данной работе, составляют
публикации  \cite{Kh34}, \cite{KhNd2007}, \cite{Khmath2009}, \cite{KhSav}, \cite{Kh2005},
\cite{Kh2006},  \cite{Kh2007}, \cite{Kh2009}, \cite{Kh2011}, \cite{khryab2012},
\cite{KhRyab_ms2012}.

Аналогичные исследования были выполнены для квадратичного гамильтониана, интегрируемость
которого доказана в \cite{sok_tsig_2002}. В частности, система $\ml$ "развалилась"\, на
три подсистемы и, таким образом, в системе с тремя степенями свободы с квадратичным
гамильтонианом выделяется уже шесть подсистем. Аналитические результаты таких
исследований будут опубликованы в отдельной работе.

Отметим, что на сегодняшней день пока так и не удалось получить алгебраического
разделения переменных ни в одной из систем ${\cal M}_k, k=1,\ldots 4$. В первую очередь
это связано с представлением систем (\ref{eq_4_0}) и (\ref{eq_4_1}) в удобном для этого
виде. В задаче о движении волчка Ковалевской в двойном силовом поле для аналогов систем
$\mo$ и $\ml$ найдено алгебраическое разделение переменных и исследована фазовая
топология \cite{KhSav}, \cite{Kh2007}, \cite{Kh2009}, \cite{Kh2011}. Опираясь, прежде
всего, на работы М.\,П.~Харламова, мы надеемся и в нашем случае на возможность явного
алгебраического разделения переменных в системах $\mo$ и $\ml$ .

Автор глубоко признателен д.ф.-м.н., профессору М.\,П.~Харламову за стимулирование и
обсуждение результатов, за настойчивое требование понимания тех топологических эффектов,
которые обнаружены в новой системе. Достаточно сказать, что в системе c тремя степенями
свободы обнаружена перестройка "pitch-fork"\,  (вы\-рож\-денная особенность ранга $2$),
как ориентируемая (неустойчивая к малым возмущениям), так  и неориентируемая (наоборот,
устойчивая). На рис.~6 перестройке типа "pitch-fork"\,  отвечают точки "$q$"\, и "$e$".
Для систем с двумя степенями свободы модельные перестройки такого типа описаны в
\cite{bolfom1999} (\cite[т.~1,~пример~2,~c.~24]{bolfom1999}). Другое наблюдение связано с
круговыми молекулами на изоэнергетических уровнях. Например, особым точкам $d$, $l$,
$c(m)$, $b$  на рис.~6, в прообразе которых одна или две компоненты, соответствуют
невырожденные особенности ранга $1$ типа "седло-седло"\,. Им отвечают круговые молекулы
особенностей ранга $0$ того же типа после умножения на топологическую окружность
\cite[теорема~9.6 (А.\,В.~Болсинов), c.~360]{bolfom1999}. Можно надеяться на получение
классификационных теорем о строении круговых молекул невырожденных особенностей ранга $1$
типа "седло-седло"\, с одной или двумя компонентами на слое на изоэнергетическом уровне в
системах с тремя степенями свободы. Аналогичные теоремы в интегрируемых системах с двумя
степенями свободы сформулированы в \cite{bolfom1999} (\cite[теорема~9.6
(А.\,В.~Болсинов), c.~360; теорема~9.8. (В.\,С.~Матвеев), c.~362]{bolfom1999}).

Работа носит далеко не законченный характер и тем не менее ясна вся картина бифуркаций
двумерных и трехмерных торов как внутри подсистем, так и  во всей системе соответственно.
Предполагается написание компьютерной программы с использованием среды "Mathematica",
которая бы детально моделировала атлас бифуркационных диаграмм, а также бифуркационные
диаграммы как самих подсистем, так и всей системы в целом.

Работа выполнена при поддержке РФФИ (грант \No~10-01-00043).

\end{document}